\documentclass[sigconf,balance=true,hidelinks,table]{acmart}

\usepackage{hyperref}
\usepackage{amsmath,amsfonts}
\usepackage{algorithmic}
\usepackage[ruled,linesnumbered,noend]{algorithm2e}
\usepackage{xcolor}
\usepackage{graphicx}
\usepackage{textcomp}
\usepackage{xcolor}
\usepackage{nccmath}
\usepackage{environ}
\usepackage{tcolorbox}
\usepackage{soul}
\usepackage{booktabs}
\usepackage{multirow}
\usepackage{array}
\usepackage{float}
\usepackage{makecell}
\usepackage{flushend}
\usepackage[inline]{enumitem}
\usepackage{hhline}
\usepackage{comment}
\usepackage{epstopdf}
\usepackage{caption}
\usepackage{subcaption}
\usepackage{diagbox}
\usepackage{xcolor}
\usepackage{listings}

\NewEnviron{myequation}{%
\begin{equation}
\scalebox{1}{$\BODY$}
\end{equation}
}

\newcommand{\fig}{Fig.}
\newcommand{\tab}{Table}

\newcolumntype{P}[1]{>{\centering\arraybackslash}p{#1}}
\newcommand{\code}[1]{\textnormal{\texttt{#1}}}
\acmConference[ESEM 2024]{The 18th ACM/IEEE International Symposium on Empirical Software Engineering and Measurement}{20–25 October, 2024}{Barcelona, Spain}

\copyrightyear{2024}
\acmYear{2024}
\setcopyright{acmlicensed}\acmConference[ESEM '24]{Proceedings of the 18th ACM / IEEE International Symposium on Empirical Software Engineering and Measurement}{October 24--25, 2024}{Barcelona, Spain}
\acmBooktitle{Proceedings of the 18th ACM / IEEE International Symposium on Empirical Software Engineering and Measurement (ESEM '24), October 24--25, 2024, Barcelona, Spain}
\acmDOI{10.1145/3674805.3686670}
\acmISBN{979-8-4007-1047-6/24/10}

\begin{document}
\title{Automated Code-centric Software Vulnerability Assessment:\\ How Far Are We? An Empirical Study in C/C++}

\author{Anh The Nguyen}
\affiliation{Independent Researcher \country{Vietnam}}
\email{anhnguyenthe.99.06.05@gmail.com}

\author{Triet Huynh Minh Le}
\affiliation{\institution{CREST - The Centre for Research on Engineering Software Technologies, The University of Adelaide}
\city{Adelaide}
\country{Australia}}
\affiliation{\institution{Cyber Security Cooperative Research Centre, Australia}
\city{}
\country{}}
\email{triet.h.le@adelaide.edu.au}

\author{M. Ali Babar}
\affiliation{\institution{CREST - The Centre for Research on Engineering Software Technologies, The University of Adelaide}
\city{Adelaide}
\country{Australia}}
\affiliation{\institution{Cyber Security Cooperative Research Centre, Australia}
\city{}
\country{}}
\email{ali.babar@adelaide.edu.au}

\begin{abstract}

\textbf{Background:}
The C/C++ languages hold significant importance in Software Engineering research because of their widespread use in practice.
Numerous studies have utilized Machine Learning (ML) and Deep Learning (DL) techniques to detect software vulnerabilities (SVs) in the source code written in these languages. 
However, the application of these techniques in function-level SV assessment has been largely unexplored. 
SV assessment is increasingly crucial as it provides detailed information on the exploitability, impacts, and severity of security defects, thereby aiding in their prioritization and remediation.
\textbf{Aims:}
We conduct the first empirical study to investigate and compare the performance of ML and DL models, many of which have been used for SV detection, for function-level SV assessment in C/C++.
\textbf{Method:}
Using 9,993 vulnerable C/C++ functions, we evaluated the performance of six multi-class ML models and five multi-class DL models for the SV assessment at the function level based on the Common Vulnerability Scoring System (CVSS).
We further explore multi-task learning, which can leverage common vulnerable code to predict all SV assessment outputs simultaneously in a single model, and compare the effectiveness and efficiency of this model type with those of the original multi-class models. 
\textbf{Results:}
We show that ML has matching or even better performance compared to the multi-class DL models for function-level SV assessment with significantly less training time.
Employing multi-task learning allows the DL models to perform significantly better, with an average of 8--22\% increase in Matthews Correlation Coefficient (MCC), than the multi-class models.
\textbf{Conclusions:}
We distill the practices of using data-driven techniques for function-level SV assessment in C/C++, including the use of multi-task DL to balance efficiency and effectiveness.
This can establish a strong foundation for future work in this area.

\end{abstract}



\maketitle
\section{Introduction}

C and C++ are among the most popular programming languages currently in use.
They are also associated with the highest number of Software Vulnerabilities (SVs) in real-world projects reported on the Common Vulnerabilities and Exposures (CVE)~\cite{bhandari2021cvefixes} database.
These SVs pose serious security threats as they can be exploited, leading to adverse consequences, such as unauthorized access to a system, loss of data, or system failure~\cite{nvd_website}. 
Given the extensive usage of C/C++ in critical systems and infrastructures, especially where low-level control of resources is required, the impact of SVs in these languages can be devastating. 
An infamous example is the Heartbleed bug that originated in OpenSSL, which was described at the time of discovery (2014) as one of the most catastrophic security risks in history~\cite{carvalho2014heartbleed,kyatam2017heartbleed}. 
The SV allowed attackers to remotely access protected memory in an estimated 55\% of all HTTPS websites~\cite {durumeric2014matter}, exposing many users' credentials and affecting a multitude of services around the world. 
As a result, many studies have focused on C and C++ for SV detection, particularly aiming to predict the presence of SVs within functions~\cite{hanif2021rise}.

Resolving SVs, a crucial step in ensuring the robustness of the system or software affected by SVs, requires more than just the information from the detection step.
Given the nature of these SVs being not equally important or urgent, they cannot be remediated all at once due to time and resource constraints~\cite{khan2018review}.
In practice, SV assessment can be employed to prioritize different SVs based on their characteristics and level of severity.
A notable framework for SV assessment is the Common Vulnerability Scoring System (CVSS)~\cite{cvss_website}, which defines a standardized set of metrics on the exploitability, impact, and severity of SVs. 
Leveraging these metrics, fixing prioritization can be set on the most critical SVs to limit their risks to the affected projects as early as possible.
Previous studies on the predictions of these CVSS metrics have proposed different data-driven techniques to automate SV assessment using SV reports (e.g., ~\mbox{\cite{han2017learning,spanos2018multi,le2019automated,le2022survey,le2022towards,le2024mitigating}}).
However, these previous studies have mostly neglected performing SV assessment directly in code functions, potentially leading to untimely fixing prioritization due to missing/delayed public reporting of SVs~\cite{li2017large,piantadosi2019fixing,le2022use}.

Data-driven approaches have been extensively used for function-level SV detection~\cite{ghaffarian2017software,lin2020deep,hanif2021rise,fu2022linevul,le2024latent,le2024software}, but these techniques have hardly been investigated for the assessment of SV, despite the similarity in the input code and the importance of the two tasks.
Specifically, there are two main gaps in the use of data-driven approaches for function-level SV assessment.
The \textit{first} gap is the lack of a performance comparison between Machine Learning (ML) and Deep Learning (DL) for function-level SV assessment, specifically in C/C++ where most SVs have been found.
For SV detection, ML was first explored with promising performance~\cite{ghaffarian2017software}; recently proposed DL models have been shown to be more effective than traditional ML approaches, achieving state-of-the-art (SOTA) performance ~\cite{zhou2019devign,le2020deep,nguyen2022regvd, chen2023diversevul,wang2020combining,li2018vuldeepecker,fu2022linevul}.
Without a similar investigation for SV assessment, it remains unknown to the practice which type of models (ML or DL) should be selected to optimize performance.
Moreover, SV assessment comprises multiple prediction tasks (e.g., exploitability, impact, and severity); therefore, building separate models for each task may induce high training and maintenance costs.
A potential solution to this is multi-task learning, i.e., building a single model to predict multiple related tasks simultaneously while sharing the same input/code. This approach has been shown to improve the predictive performance of many Software Engineering tasks~\cite{gong2019joint, liu2023interpretable, li2023vulanalyzer, mastropaolo2021studying, shetty2021neural, wang2024mtl,liu2020multi, liu2020self,deng2024mtlink}.
However, the effectiveness of multi-task learning for function-level SV assessment still remains unknown, posing the \textit{second} gap.

We conduct this empirical study to fill the two aformentioned research gaps.
Using a customized dataset with vulnerable C/C++ functions and their respective CVSS metrics, we study the performance of various ML and DL models to automate function-level SV assessment in C/C++.
The prediction outputs are the characteristics of these vulnerable functions based on the CVSS metrics. We evaluate and compare both the multi-class ML and DL models, as well as the multi-learning variants of DL models.
Our findings indicate that ML performs competitively for the SV assessment tasks with an average of the best model, achieving an effectiveness value of 0.680 in Matthews Correlation Coefficient (MCC).
Surprisingly, the best ML model outperforms all multi-class DL models in six out of seven CVSS metrics, while requiring only one-fourth of the total training time.
Switching from multi-class to multi-task learning allows the DL models to obtain significantly better MCC, with an average increase of 8--22\% in MCC compared to the original performance. The best multi-task DL model also has an 8\% higher MCC value than the best ML model.
Our key \textbf{contributions} are:
\begin{itemize}[noitemsep,topsep=0pt]
    \item We are the first to investigate data-driven approaches for function-level SV assessment in C/C++, the languages with the most reported SVs.
    \item We distill the practices of using ML and DL for function-level SV assessment. Multi-class ML is better suited for resource-limited scenarios with comparable performance to multi-class DL and significantly less training time. Conversely, multi-task DL is the optimal choice for the highest performance.
    \item We release our data, code, and models for future research~\cite{reproduction_package_dl4sa}.
\end{itemize}

\noindent Overall, our study provides knowledge-based evidence and recommended practices for both researchers and practitioners on how to select and train data-driven models to enable effective and efficient assessment of SVs at the function level.

\noindent \textbf{Paper structure}.
Section~\mbox{\ref{sec:background}} introduces the background of CVSS-based SV assessment in C/C++ functions.
Section~\mbox{\ref{sec:rqs}} presents the two RQs.
Section~\mbox{\ref{sec:method}} describes the methods employed to answer these RQs. Section~\mbox{\ref{sec:results}} distills our results. 
Section~\mbox{\ref{sec:discussion}} discusses these findings and threats to validity.
Section~\mbox{\ref{sec:conclusions}} concludes the study.
\vskip 20pt

\section{Related Work and Motivation}

\begin{table*}[t]
\resizebox{\textwidth}{!}{%
\begin{tabular}{l|l|l}
\hline
\textbf{Metric group} & \textbf{CVSS metric} & \textbf{Description} \\ \hline
\multicolumn{1}{c|}{\multirow{3}{*}{Exploitability}} & Access Vector & Medium/technique to   attack/penetrate a system \\ \cline{2-3} 
\multicolumn{1}{c|}{} & Access Complexity & The complexity to exploit the SV and initiate an attack \\ \cline{2-3} 
\multicolumn{1}{c|}{} & Authentication & Whether/what authentication the attack requires to happen \\ \hline
\multirow{3}{*}{Impact} & Confidentiality & The extent to which the attack allows unauthorized access to system sensitive/confidential data \\ \cline{2-3} 
 & Integrity & The extent to which the attack allows unauthorized modifications to system data \\ \cline{2-3} 
 & Availability & The extent to which the attack restricts accessibility to system data, resources, and services \\ \hline
Severity & Severity Level & Combination of the Exploitability and Impact metrics, approximating the criticality of SVs \\ \hline
\end{tabular}%
}
\caption{The CVSS assessment metrics considered in this study and their descriptions.}
\label{tab:cvss_metrics}
\vspace{-13pt}
\end{table*}

\label{sec:background}
\subsection{SV Assessment with CVSS}
\label{subsec:cvss_metrics}

SV assessment plays a crucial role in the SV lifecycle, defining different attributes of identified SVs~\cite{smyth2017software}.
These attributes aid developers in comprehending the characteristics of SV, thus guiding prioritization and remediation approaches.
For instance, an SV capable of significantly compromising system confidentiality, such as enabling unauthorized access to sensitive data, warrants high-priority resolution.
Subsequently, a protocol can be implemented to safeguard confidentiality, like verifying or enforcing access privileges for the impacted component or data.

The Common Vulnerability Scoring System (CVSS)~\cite{cvss_website} is a widely employed framework for SV assessment by researchers and practitioners.
It comprises two main versions: versions 2 and 3, with version 3 introduced in 2015.
Despite this, CVSS version 2 (CVSSv2) remains prevalent because many SVs predating 2015 still pose risks to modern systems.
For instance, an SV identified as CVE-2004-0113, originating in 2004, was exploited in 2018~\cite{old_sv_exploit}.
Therefore, we utilize the assessment metrics of CVSS version 2 as the outputs for SV assessment models in our study.

Although CVSSv2 defines three groups of metrics: Base, Temporal, and Environmental, we intended to only capture the nature characteristics of the vulnerabilities. Therefore, we only selected the metrics which are mandatory for the calculation of the score (Severity) from the \textit{Base} group (Access Vector, Access Complexity, Authentication, Confidentiality, Integrity, Availability).
These metrics and their descriptions are given in \tab~\ref{tab:cvss_metrics}.
The other two groups (\textit{Temporal} and \textit{Environmental}), while providing valuable context information about SV changes and their environments, are optional in the score calculation and only exists in some CVE records; thus, we decide to exclude them.

Additionally, while Severity is a combination of Exploitability and Impact, it does not necessarily provide a holistic view of SVs, potentially leading to a sub-optimal SV fixing plan.
For example, according to the CVSS specification~\cite{cvss_v2}, two SVs would have the same severity level if they share the same exploitability but affect different attributes (e.g., Confidentiality vs. Integrity) of a system to the same extent.
As a result, to ensure a thorough assessment of SVs, we utilize all of the seven mentioned metrics (i.e., Confidentiality, Integrity, Availability, Access Vector, Access Complexity, Authentication, and Severity) as the outputs for building SV assessment, akin to prior studies (e.g.,~\cite{spanos2018multi,le2019automated,gong2019joint,le2022use}).

\captionsetup[lstlisting]{labelsep=period, textfont=footnotesize, labelfont=footnotesize, justification=justified}

\newcommand{\lstbg}[3][0pt]{{\fboxsep#1\colorbox{#2}{\strut #3}}}
\lstdefinelanguage{diff}{
  basicstyle=\ttfamily\small,
  morecomment=[f][\lstbg{red!20}]-,
  morecomment=[f][\lstbg{green!20}]+,
}

\definecolor{difftitle}{HTML}{000099}
\definecolor{diffstart}{HTML}{660099}
\definecolor{diffincl}{HTML}{006600}
\definecolor{diffrem}{HTML}{AA3300}

\definecolor{del_color}{HTML}{FFCCCC}
\definecolor{add_color}{HTML}{CCFFCC}

\lstdefinestyle{lst}{
    numbers=left, 
    numberstyle=\scriptsize, 
    numbersep = 5pt,
    framexleftmargin = 0in,
    framexrightmargin = 0in,
    xleftmargin = 0.18in,
    xrightmargin = 0.1in,
    basicstyle=\ttfamily\scriptsize, 
    frame=lines,
    showtabs=true,
    showspaces=true,
    showstringspaces=false,
    literate={\ }{{\ }}1,
    escapeinside={<@}{@>}
}

\lstset{belowskip=-0.05in}

\begin{figure}

\begin{lstlisting}[language=diff,style=lst]
char *strdup(const char *s1)
{
    char *s2 = 0;
    if (s1) {
-       s2 = malloc(strlen(s1) + 1);
-       strcpy(s2, s1);
+       size_t len = strlen(s1) + 1;
+       s2 = malloc(len);
+       memcpy(s2, s1, len);
    }
    return s2;
}
\end{lstlisting}
\caption{An example of a C++ vulnerable function in \textit{Git}, extracted from the fixing commit \textit{34fa79a} of the SV [\textit{CVE-2016-2315}]. Note: line 5 and 6 are vulnerable. Additions and removals are colored in green and red, respectively.}
\label{fig:example_sv_assessment}
\vspace*{-\baselineskip}
\end{figure}

\subsection{Data-Driven SV Assessment: From SV Reports to Function Source Code}
\label{subsec:cvss_metrics}
After SVs are detected, developers need to investigate the root cause, design the fixing or mitigation plan, implement these fixes, and roll out these changes after testing; all of which require great level of effort.
Therefore, automating SV assessment is an important step since determining the severity and scale of impact can help inform decisions, prioritization, and planning of resources to be made accordingly. 
Previous studies in the automation of SV assessment have utilized SV reports as the predominant form of input for the predictions (e.g.,~\mbox{\cite{han2017learning,spanos2018multi,le2019automated,duan2021automated,le2022survey}}). 
However, such an approach is inherently constrained by the availability of these reports.
Publishing SV reports after the SVs are fixed is a recommended practice to mitigate their exploitation \cite{le2021large,zhou2021finding}. 
For instance, \textit{CVE-2016-2315}, an extremely critical SV with a reported CVSSv2 severity score of 10.0/10.0, was fixed 185 days prior\footnote{https://github.com/git/git/commit/34fa79a6cde56d6d428ab0d3160cb094ebad3305} to its public disclosure on August 4th, 2016.\footnote{https://nvd.nist.gov/vuln/detail/CVE-2016-2315} 
This practice limits the application of report-level inputs to prioritize SVs in time.

Taking inspiration from function-level SV detection, using vulnerable functions directly can enable SV assessment prior to fixing, reducing the above-mentioned delay caused by report-level SV assessment.
To illustrate function-level SV assessment, we use the function in ~\fig~\ref{fig:example_sv_assessment}, one vulnerable function extracted from the SV \textit{CVE-2016-2315} discussed earlier, which exhibits a possible buffer overflow error during the handling of long path names. 
The original code used the built-in \code{strcpy} function, which did not take into account the length of the \code{s1} argument (the source string to copy from), but instead copied from the start to a \code{null} character. 
If an attacker crafts a special string without a \code{null} character, the function will copy beyond the end of the buffer and can trigger this SV.
The attacker can feed the malicious input to the system from any network remotely, so the Access Vector is classified as Network and the Access Complexity is Low.
The attacker does not require authentication to either the actual server/computer running the code, so its Authentication is None. 
The malicious input, running inside the file system of affected machines, can leak full information about the machines and connected devices, bypassing all security protocols and possibly disabling all affected resources completely.
This resulted in this SV being categorized as Complete for Confidentiality, Integrity, and Availability impacts.
All of these factors render this SV one of the most severe SVs with the highest possible score of 10.0 in the Severity category of CVSSv2.
The fixing commit replaced all occurrences of \code{strcpy} with \code{memcpy}, which copies an exact amount of bytes provided with a length argument, removing the possibility of an overflow even if the string itself does not contain a null character.
Although the fix is simple in concept, without the assessment information, developers can overlook the SV and treat this as a low-level threat, thus motivating the need to automate SV assessment at the function level.

While Le et al.~\cite{le2022use}'s study is the most related to our work, ours is fundamentally different from theirs.
First, the previous investigation was conducted in Java, which does not have as many critical SVs as in C/C++, limiting the significance of the work in relation to the latest SV-related literature and real software development environments.
Moreover, they did not study any use of Deep Learning (DL) for SV assessment, which does not provide a thorough understanding of \textit{how far we are} with function-level SV assessment, given that DL is the state-of-the-art for SV detection.

\subsection{Deep Learning: SOTA for SV Detection Yet Under-Explored for SV Assessment}
\label{subsec:dl_sv_assessment}

With the rise of data-driven methods in recent years, DL has also been applied more commonly to SV-related tasks~\cite{harzevili2023survey}. 
In particular, the application of DL in detecting SVs in source code at the function level has been explored extensively, with multiple works using different approaches, including sequential-based, graph-based, and Transformer-based models.
An overview of these models can be found in \tab~\ref{tab:dl_sv_detection}. 
It should be noted that, given the extensive array of studies, this list does not aim to cover all of the proposed models, but it rather shows the key architectures having been used for function-level SV detection.

Sequence-based DL models, particularly those implemented with convolutional or recurrent neural networks, have been widely applied to SV detection (e.g., \cite{filus2021random, hoang2019deepjit, li2017software, dam2018automatic, liu2019deepbalance}) due to their advantages over traditional ML methods. 
\textit{Convolutional Neural Network} (CNN) \cite{kim2014convolutional} have been proven capable of identifying local patterns and structures within source code \cite{russell2018automated} and have been used widely by recent studies as a baseline for newer models to detect SVs at the function level.
Recurrent models are also popular, accounting for 28\% of primary studies in SV detection at different levels of granularity~\cite{harzevili2023survey}. Among the models that adopted recurrent architecture, \textit{Long Short-term Memory} (LSTM) \cite{hochreiter1997long} is particularly suitable for discerning the underlying sequential order of source code, especially C/C++ source code \cite{karpathy2015visualizing}, by keeping control of the information from one step/token to the next~\cite{hochreiter1997long}.

Another effective and popular architecture is the graph-based models, which specifically deal with graph-like structured data. 
Graphs, representations of inputs as nodes linked by edges, are ideal for capturing relationships and dependencies. 
Therefore, graph models are generally more powerful than recurrent or convolutional models due to their ability to capture more meaningful context, especially the semantic representation of complex structures in source code~\cite{khemani2024review}. 
Since SVs originate from source code, graph-based models are expected to be well-suited for recognizing the underlying structural information. Among these, Graph Convolutional Networks (GCN) \cite{kipf2016semi} and Gated Graph Neural Networks (GGNN) \cite{li2015gated} have seen the most usage and better performance than the sequence-based models.
GCN is similar to traditional CNN, where the nodes in a graph are learned in the context of the neighbor nodes through the use of Graph convolution layers, linear layers, and activation functions~\cite{kipf2016semi}.
GGNN, on the other hand, leverages Gated Recurrent Unit \cite{cho2014learning} to keep the retrieved information in recurrent steps, and gradient is calculated by backpropagation.

Given the widespread popularity of Bidirectional Encoder Representations from Transformers (BERT) \cite{devlin2018bert} in the field of Natural Language Processing, models based on the Transformer architecture~\cite{vaswani2017attention}, especially CodeBERT~\cite{feng2020codebert}, have been utilized to perform SV detection~\cite{fu2022linevul}. 
CodeBERT uses multi-layer bidirectional Transformer \cite{vaswani2017attention}, and is pre-trained on both natural language and code datasets. 
This allows CodeBERT to achieve SOTA performance in code-related downstream tasks. Fu et al.~\cite{fu2022linevul} fine-tuned CodeBERT on a dataset of C/C++ SVs to develop LineVul. 
This model has been shown to outperform existing sequential-based and graph-based models and achieve the SOTA performance for function-level SV.

To improve effectiveness and efficiency, multi-task deep learning performs the prediction of multiple related tasks in a single model, which is applicable to function-level SV assessment tasks.
Multi-task learning is the practice of incorporating all training tasks and data into one model and performing training of all the constituent tasks simultaneously.
By utilizing data from different tasks and the knowledge-sharing ability of DL models, multi-task DL models can learn more universal representations of inputs, thus producing a superior performance and reducing the overfitting likelihood for each task~\cite{zhang2021survey}. 
This approach has also been successfully adopted in a variety of code-based tasks, such as source code understanding \cite{liu2020multi}, code completion \cite{liu2020self}, issue-commits linking \cite{deng2024mtlink}.
These tasks utilize functions source code as inputs, and their performance improved noticeably when compared to the traditional multi-class ML/DL approaches.

While these studies have showcased the potential of DL and multi-task DL in various code-based Software Engineering tasks, particularly SV detection, their application to function-level SV assessment is still limited. Therefore, to the best of our knowledge, our empirical study is the first to (1) compare the performance of state-of-the-art DL and ML models for SV assessment and (2) investigate whether multi-task learning can enable more effective and efficient predictions of the tasks.
\begin{table}[t]
\fontsize{8}{9}\selectfont
\resizebox{\columnwidth}{!}{%
\begin{tabular}{lll}
\textbf{Architecture}          & \textbf{Model}          & \textbf{References}                                                 \\ \hline
\multirow{5}{*}{Sequence-based}        & \multirow{2}{*}{CNN}            & Li et al. 2017 \cite{li2017software}               \\
                                       &                                 & Yan et al. 2021 \cite{yan2021han}                  \\ \cline{2-3} 
                                       & \multirow{5}{*}{RNN}           & Dam et al. 2017 \cite{dam2017automatic}            \\
                                       &                                 & Li et al. 2018 \cite{li2018vuldeepecker}            \\
                                       &                                 & Liu et al. 2020 \cite{liu2020cd}                   \\
                                       &                                 & Li et al. 2021 \cite{li2021sysevr}            \\
                                       &                                 & Zou et al. 2021 \cite{zou2021interpreting}         \\ \hline
\multirow{5}{*}{Graph-based}    & \multirow{3}{*}{GGNN}   & Zhou et al. 2019 \cite{zhou2019devign}   \\
                                &                         & Chakraborty et al. 2020 \cite{chakraborty2021deep}   \\
                                & & Ding et al. 2022 \cite{ding2022velvet}             \\ 
                                & & Nguyen et al. 2022 \cite{nguyen2022regvd}               \\ \cline{2-3} 
                                & \multirow{2}{*}{GCN}    & Cheng et al. 2021 \cite{cheng2021deepwukong}       \\
                                &                         & Ghaffarian et al. 2021 \cite{ghaffarian2021neural} \\ & & Nguyen et al. 2022 \cite{nguyen2022regvd}               \\ \hline
\multirow{2}{*}{Transformer-based} & \multirow{2}{*}{CodeBERT} & Feng et al. 2020 \cite{feng2020codebert} \\
                               &                         & Fu et al. 2022 \cite{fu2022linevul} \\ \hline      
\end{tabular}
}
\caption{DL models used in SV detection. Note: CNN: Convolutional Neural Network, RNN: Recurrent Neural Network, GGNN: Gated Graph Neural Network, GCN: Graph Convolutional Network.}
\label{tab:dl_sv_detection}
\vspace{-30pt}
\end{table}

\begin{figure*}[t]
    \centering
    \includegraphics[trim={18.5cm 7.5cm 19cm 3cm},clip,width=\textwidth,keepaspectratio]{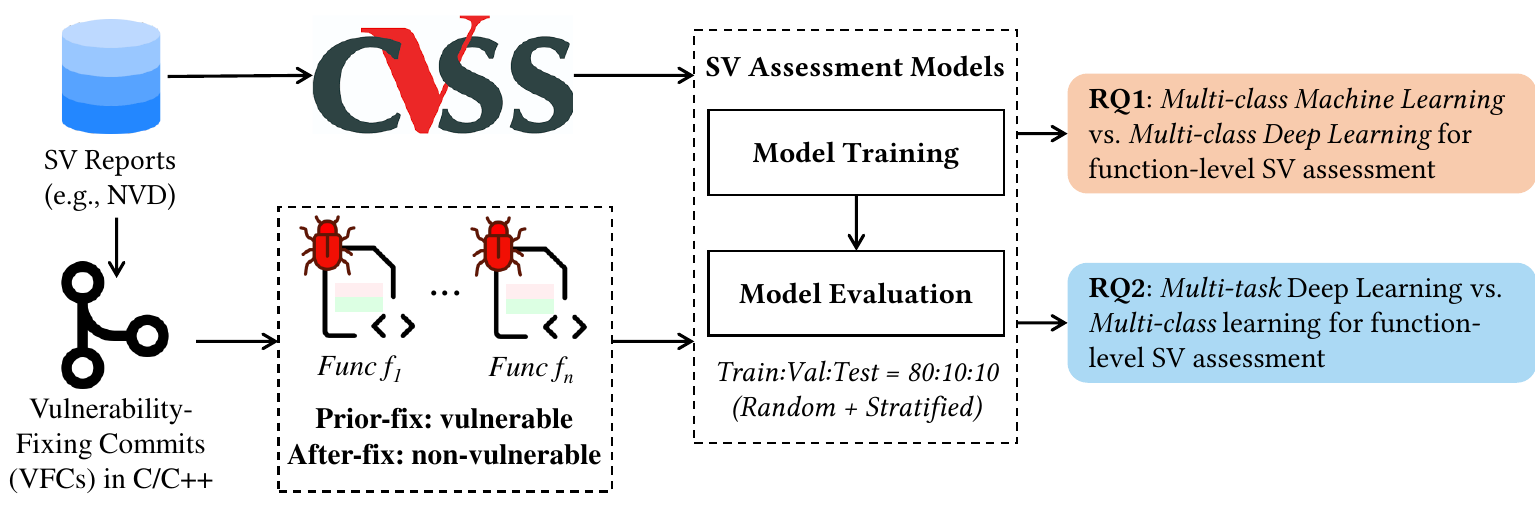}
\caption{Methodology used to answer the research questions.}
\label{fig:workflow}
\vspace{-10pt}
\end{figure*}

\section{Research Questions}
\label{sec:rqs}

To explore and compare the performance of ML and DL models for assessing SVs with CVSS metrics in the C/C++ languages, we investigate the following two Research Questions (RQs).

\textbf{RQ1: How well do multi-class Machine Learning and Deep Learning models perform SV assessment in C/C++?}
It is possible to automate the prediction of CVSS assessment metrics using ML models with vulnerable code as input~\cite{le2022use}. However, the prior investigation was not done on C/C++, the most commonly investigated language in the area of SV detection, the prior step to SV assessment (see section~\ref{sec:background}).
This presents a gap between the current practice of SV detection and SV assessment.
Moreover, in recent years, DL models have achieved SOTA performance for SV detection~\cite{lin2020software}. However, these DL models have never been utilized for function-level SV assessment.
Thus, RQ1 aims to evaluate and compare the performance of the recently proposed ML models and the representative DL models, that have been previously used for SV detection, for SV assessment in C/C++.
RQ1 also gives insights into the optimal feature/classifier choices for the tasks, enabling effective function-level SV assessment.

\textbf{RQ2: Can multi-task Deep Learning improve SV assessment in C++?}
While RQ1 focuses on multi-class models, i.e., building a separate model for each SV assessment task, RQ2 aims to study the possibility of using multi-task learning, i.e., combining all the tasks in a single model.
It is worth noting that multi-task learning is mainly targeted to DL models, so we will adapt the DL models used in RQ2 for this setting.
More details are given in section~\ref{subsec:assessment_models}.
RQ2 also compares the performance of multi-task models with that of the multi-class models presented in RQ1, giving insights into whether/to what extent DL can improve upon ML for function-level SV assessment. 
RQ2 findings can inform whether the function-level SV assessment tasks would benefit from sharing input features through multi-task learning to support efficient model building.

\section{Research Methodology}
\label{sec:method}

This section presents the methodology we used to empirically study the use of ML and DL for function-level SV assessment to support timely prioritization and fixing of SVs in C/C++. We used a machine with 8 CPU cores, 16GB of RAM, and GeForce RTX 3050Ti GPUs to carry out all the experiments.

\noindent \textbf{Overview}. \fig~\ref{fig:workflow} depicts the workflow we used to answer the two RQs given in section ~\ref{sec:rqs}. The workflow consists of three main steps: (\textit{i}) data collection, (\textit{ii}) model building and evaluation, and (\textit{iii}) results reporting.
First, we curated a dataset of SVs in real-world projects written in C/C++. We then extracted the vulnerable code functions modified in the vulnerability-fixing commits of these reported SVs as input as well as the seven CVSS metrics as output. More details of the data collection are given in section~\ref{subsec:data_collection}.
Secondly, the extracted input and output were used to develop multi-class ML and multi-class DL models in RQ1 and customized to support multi-task learning in RQ2.
The details of these models building and evaluation are given in sections~\ref{subsec:assessment_models} and~\ref{subsec:model_evaluation}. Finally, we evaluated and reported the performance of ML and DL in RQ1 as well as those of multi-task models and multi-class models in RQ2.

\subsection{Data Collection}
\label{subsec:data_collection}

\begin{figure}[t]
    \centering
    \includegraphics[width=\columnwidth,keepaspectratio]{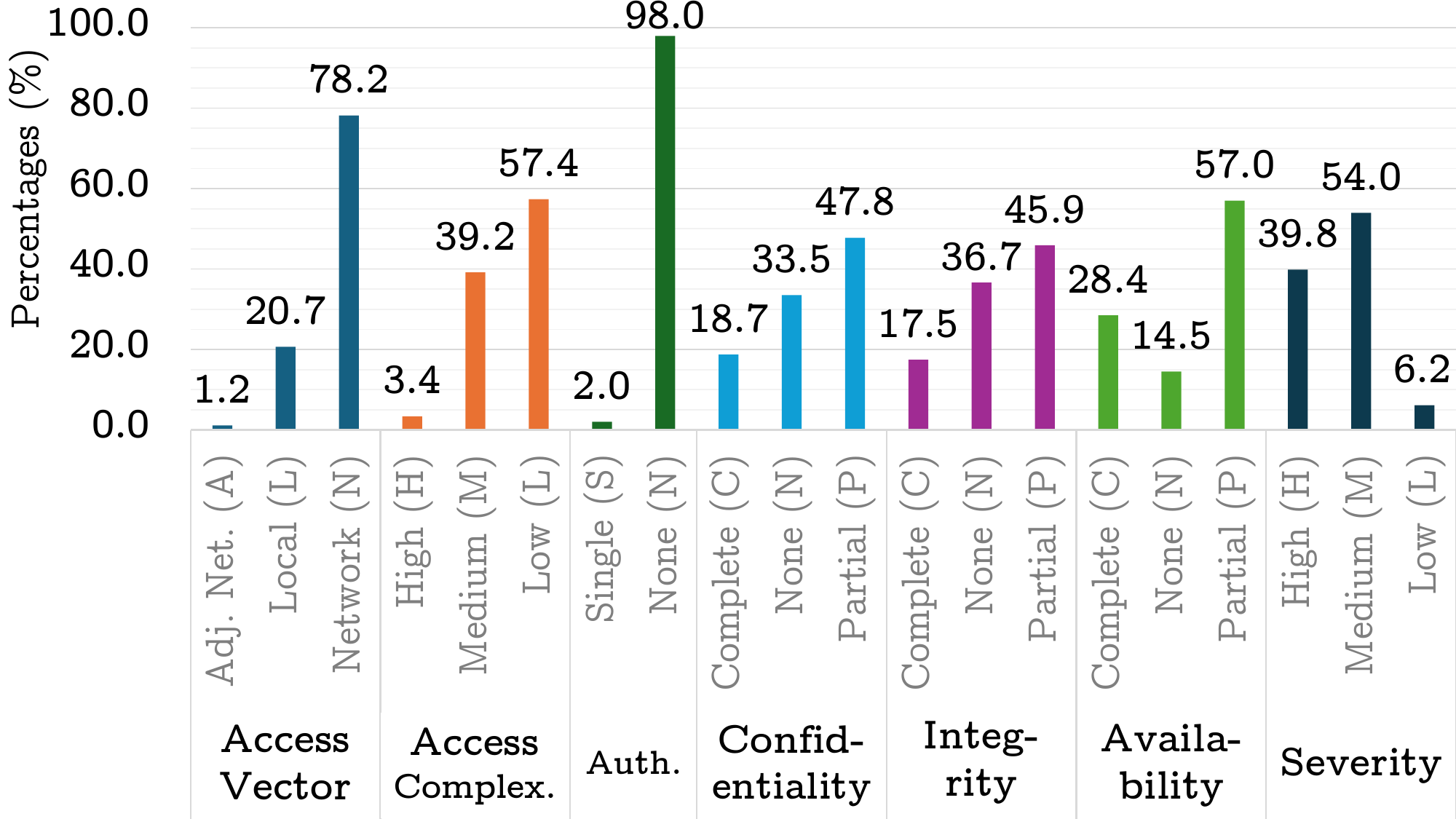}
\caption{The CVSS metrics distributions in the dataset. Note: Access Complex., Auth., and Adj. Net. refer to Access Complexity, Authentication, and Adjacent Network, respectively.}
\label{fig:cvss_distribution}
\vspace{-16pt}
\end{figure}

\noindent \textbf{Dataset selection}. Developing models to assess SV requires a specially curated dataset that contains vulnerable functions and their corresponding CVSS metrics. We built our dataset upon \textit{Big-Vul}~\cite{fan2020ac} as it is one of the most popular published SV datasets in C/C++ due to its size and rich features. Big-Vul consisted of 188,636 functions written in C/C++ related to 91 different SV types originating from 348 different projects on GitHub.
These functions were extracted from respective vulnerability-fixing commits of the SVs from public CVE databases spanning from 2002 to 2019.
In this dataset, vulnerable functions were those with lines altered in the gathered VFCs, whereas those that remained unchanged were labeled as non-vulnerable. The dataset provided 21 features for each function, including the CVE-ID, the project name of the source code, and the fixed version of each function.
It is worth noting that there are other SV datasets, such as the FFmpeg-Qemu dataset in Devign~\cite{zhou2019devign}, the Chromium-Debian dataset in ReVeal~\cite{chakraborty2021deep}, or DiverseVul~\cite{chen2023diversevul}, but we did not consider them as they did not include the CVSS metrics and/or CVE-ID required to trace the CVSS metrics.
The customization we made to the Big-Vul dataset to suit SV assessment are described hereafter.

\noindent \textbf{Data preparation}. We focused on the assessment of detected SVs, so we filter for the vulnerable functions in the Big-Vul dataset.
We also wanted to ensure the relevance of the data being used, and thus we checked whether each of the original functions was still available on GitHub.
Consequently, we discarded the functions for which we could no longer confirm their existence at the time of data collection in October 2023.
For each of the selected functions, we retrieved the seven CVSSv2 metrics given in section~\ref{subsec:cvss_metrics}, and noticed that Access Vector was missing in the Big-Vul dataset.
Thus, we relied on the CVE-ID of each vulnerable function to retrieve the value of Access Vector from public CVE databases.
In addition, we converted the base scores to the severity levels as follows: 0.1 - 3.9 for Low, 4.0 - 6.9 for Medium, and 7.0 - 10.0 for High, following the NVD CVSS qualitative ratings \cite{nvd_cvss_rating}.
After the data collection process, we obtained the final dataset with \textit{9,993} records, each consisting of a vulnerable function and the respective seven CVSSv2 metrics, including Access Vector, Access Complexity, Authentication, Confidentiality, Integrity, Availability and Severity. 
The distributions of the curated dataset in terms of CVSS metrics are given in~\fig~\ref{fig:cvss_distribution}.

\subsection{Machine Learning and Deep Learning Models for Function-Level SV Assessment}
\label{subsec:assessment_models}
This section describes the implementation of the models. In total, we utilized six ML models and five DL models.
These models were modified to enable multi-class classification of the CVSS metrics if they had not been done so originally.
We also customized the DL models to support multi-task learning in RQ2. 
Note that we introduced multi-task learning only for DL due to the flexibility of the architecture, e.g., allowing parameter sharing in the network.
We then used the same evaluation method across all models and obtained the evaluation measures to analyze and report the results.

\subsubsection{\textbf{Machine Learning models}}\hfill
\label{subsec:assessment_ml_models}\\
ML models have previously been used extensively for SV-related tasks (e.g.,~\cite{ghaffarian2017software, spanos2018multi,le2022use}).
Specifically, we utilized three non-ensemble models: Logistic Regression (LR)~\cite{walker1967estimation}, Support Vector Machine (SVM)~\cite{cortes1995support}, K-Nearest Neighbor (KNN)~\cite{altman1992introduction}; and three ensemble models: Random Forest (RF)~\cite{ho1995random}, Extreme Gradient Boosting (XGB)~\cite{chen2016xgboost}, and Light Gradient Boosting Machine (LGBM)~\cite{ke2017lightgbm}, based on the implementations provided by Le et al.~\cite{le2020puminer,le2022use}.
We optimized each model with the commonly used hyperparameters listed in \tab~\ref{tab:hyperparameter_ml_models}.
These models used input data from four different feature extraction methods, namely \textit{Bag-of-Tokens}, \textit{Bag-of-Subtokens}, \textit{Word2Vec}, and \textit{fastText}.
\textit{Bag-of-Tokens} (BoT) originates from Bag-of-Words, which is a popular feature extraction technique in the field of Natural Language Processing. In this scenario, the frequency of each token was accumulated after a process of customized tokenization where the syntax and semantics of the code were retained.
\textit{Bag-of-Subtokens} (BoST) expands upon Bag-of-Tokens, where code tokens were further divided into character sequences known as sub-tokens. The length of the sub-tokens in our study ranged from two to six characters.
\textit{Word2Vec}~\cite{mikolov2013distributed} captures the features of any given token taking into account its neighboring tokens, allowing groups of similar tokens to generate similar feature vectors.
This improved upon the previous two techniques, BoT and BoST, by incorporating code contexts.
For this approach, the vector sizes of \textit{150, 300, 500} with the window sizes of \textit{3, 4, 5} were chosen.
\textit{fastText}~\cite{bojanowski2017enriching} extends Word2Vec in a similar fashion to Bag-of-Subtokens extending Bag-of-Words, where each token was divided into its sub-tokens, while retaining the context-capturing ability of Word2Vec. 
The final feature vectors of each code token were then aggregated from the feature vectors of all sub-tokens.
The settings for fastText we used were a combination of sub-tokens lengths from Bag-of-Subtokens and the vector sizes and window sizes of Word2Vec.

\begin{table}[t]
\resizebox{\columnwidth}{!}{%
\begin{tabular}{cll}
\textbf{Model} & \multicolumn{1}{c}{\textbf{Hyperparameters}} & \multicolumn{1}{c}{\textbf{Range / Options}} \\ \hline
Logistic Regression & \multirow{2}{*}{\begin{tabular}[c]{@{}l@{}}Regularization\\  coefficient\end{tabular}} & \multirow{2}{*}{0.01, 0.1, 1, 10, 100} \\
Support Vector Machine &  &  \\ \hline
\multirow{3}{*}{K-Nearest Neighbors} & No. of neighbors & 5, 11, 31, 51 \\
 & Weights & uniform, distance \\
 & Distance Norm & 1, 2 \\ \hline
\multirow{3}{*}{\begin{tabular}[c]{@{}c@{}}Random Forest\\  Extreme Gradient Boosting\\  Light Gradient Boosting Machine\end{tabular}} & No. of estimators & 100, 200, 300, 400, 500 \\
 & Max depth & 3, 5, 7, 9, unlimited \\
 & Max no. of leaf nodes & 100, 200, 300, unlimited \\ \hline
\end{tabular}%
}
\caption{Hyperparameters tuning for ML models.}
\label{tab:hyperparameter_ml_models}%
\vspace{-15pt}
\end{table}

\begin{table}[t]
 \resizebox{\columnwidth}{!}{%
\begin{tabular}{ccll}
\textbf{Type} & \textbf{Model} & \multicolumn{1}{c}{\textbf{Hyperparameters}} & \multicolumn{1}{c}{\textbf{Range / Options}} \\ \hline
 \multirow{-1.5}{*}{\rotatebox{90}{Graph}} & \begin{tabular}[c]{@{}c@{}}GGNN\\ GCN \end{tabular} & \begin{tabular}[c]{@{}l@{}}Learning Rate\\  Adam Optimizer Epsilon\\  Hidden size\\  No. of GNN layers\end{tabular} & \begin{tabular}[c]{@{}l@{}}5e-4 to 1e-1\\   1e-8 to 1e-4\\   32, 64, 128, 256, 512\\   1, 2, 3, 4, 5\end{tabular} \\ \hline
 & CNN & \begin{tabular}[c]{@{}l@{}}Learning Rate\\  Adam Optimizer Epsilon \\ Kernel Size\\  Padding\end{tabular} & \begin{tabular}[c]{@{}l@{}}5e-4 to 1e-1\\ 1e-8 to 1e-4\\ 1, 3, 5, 7, 9\\   0, 1, 2, 3\end{tabular} \\ \cline{2-4} 
 \multirow{-4}{*}{\rotatebox{90}{Non-graph}} & LSTM & \begin{tabular}[c]{@{}l@{}}Learning Rate\\ Adam Optimizer Epsilon \\ No. of LSTM layers\end{tabular} & \begin{tabular}[c]{@{}l@{}}5e-4 to 1e-1\\ 1e-8 to 1e-4\\ 1, 2, 3\end{tabular}\\ \cline{2-4} 
 & CodeBERT & \begin{tabular}[c]{@{}l@{}}Learning Rate\\  Adam Optimizer Epsilon\end{tabular} & \begin{tabular}[c]{@{}l@{}}5e-4 to 1e-1\\   1e-8 to 1e-4\end{tabular} \\ \hline
\end{tabular}%
}
\caption{Hyperparameters tuning for DL models. Note: Non-graph models include sequence-based models (CNN and LSTM) and Transformer-based model (CodeBERT).}
\label{tab:hyperparameter_dl_models}%
\vspace{-25pt}
\end{table}

\subsubsection{\textbf{Deep Learning models}}\hfill
\label{subsec:assessment_dl_models}\\
Similar to ML, multiple DL models have been developed for SV detection~\cite{zhou2019devign, le2021deepcva, nguyen2022regvd}. In our implementations, we mainly focused on the three key DL architectures/types that have been widely used in the literature for function-level SV assessment, including sequence-based, graph-based, and Transformer-based models.

\noindent \textbf{Sequence-based models}. For this category, we included \textit{Convolutional Neural Network (CNN)} \cite{kim2014convolutional}, \textit{Long Short-term Memory (LSTM)} \cite{hochreiter1997long}. \textit{CNN} and \textit{LSTM} are standard models for text classification tasks and can provide a robust baseline for measuring performance. 
The input data was fed into an embedding layer, then processed by either a convolution layer for CNN or an LSTM layer for LSTM, followed by a fully-connected linear layer and a softmax function for classification of the CVSS metrics.

\noindent\textbf{Graph-based models}. As shown in section~\ref{subsec:dl_sv_assessment}, there have been a number of different graph-based deep architectures proposed for SV detection, most notably Gated Graph Neural Networks (GGNN)~\cite{li2015gated} and Graph Convolutional Networks (GCN)~\cite{kipf2016semi}. To assess the performance of these representative graph-based architectures for function-level SV assessment tasks, we leveraged the publicly available implementations of Nguyen et al.~\cite{nguyen2022regvd}. These implementations have augmented the graph neural networks with different sum and max poolings to produce embedding graph features before feeding this to a fully-connected layer and softmax layer for classification.
With such augmentations, they have been demonstrated to outperform existing ones using the same architectures like Devign~\cite{zhou2019devign} and ReVeaL~\cite{chakraborty2021deep}.

\noindent\textbf{Transformer-based model}.
\textit{CodeBERT} is a popular pre-trained model for programming language and natural language that utilizes the Transformer architecture. The key advantage of CodeBERT is the ability to capture more context in the source code by producing different embedding vectors of the same tokens based on the surrounding inputs and can retain the sub-tokens that appear more frequently in the corpus via Byte-Pair Encoding (BPE) \cite{sennrich2015neural}. This has been proven to be more effective in dealing with source code input \cite{feng2020codebert}. Notably, CodeBERT has achieved the state-of-the-art predictive performance for function-level SV detection~\cite{fu2022linevul,steenhoek2023empirical}.

\noindent \textbf{Implementation of multi-class DL models for function-level SV assessment}. To assess each CVSS metric, we built each DL model with its original architecture and respective SV data (function source code and corresponding CVSS metric in the curated dataset from section~\ref{subsec:data_collection}) as input.
All of the models were also tuned with a set of hyperparameters given in \tab~\ref{tab:hyperparameter_dl_models} to find the best-performing combination for each task.

\noindent \textbf{Customization of DL models to support multi-task learning}. Instead of having seven models to perform seven SV assessment tasks as in RQ1, in RQ2, we employed multi-task learning to perform all the tasks at once, leveraging the parameter-sharing ability of DL models~\cite{zhang2021survey}.
In each DL model, we modified the last layer to include a classification head, consisting of seven linear layers, each representing a CVSS assessment task. The input for a multi-task model was the function source code and all seven CVSS metrics, and the output vector for a specific task \(i\) was defined as:
\[task_{i} = \omega_{t} x_{function} + b_{t}\]
where \(task_{i}\) is the output vector, \(\omega_{t}\) is the learnable weights, \(x_{function}\) is the feature vectors, and \(b_{t}\) is the learnable bias.

\noindent To compare this output with ground-truth labels, we then defined a loss function that is the average/mean of all the cross-entropy losses of the tasks, defined as:
\[loss = mean(\{l_1,\dots,l_N\}) \]
where \(N\) is the number of tasks from 1 to 7, and the loss for each task \(l_{n}\) is defined as:
\[l_n = - w_n \left[ y_n \cdot \log \sigma(x_n) + (1 - y_n) \cdot \log (1 - \sigma(x_n)) \right]\]
where for a given task \(n\), \(\omega_{n}\) is the weight of the loss with respect to all loss, \(y_{n}\) is the ground-truth target, and \(x_{n}\) is the predicted value.
\noindent This multi-task loss was minimized using a stochastic gradient descent method.
These multi-task models were also tuned with the same hyperparameters as their multi-class counterparts, and their hyperparameters are listed in \tab~\ref{tab:hyperparameter_dl_models}.

\subsection{Model Evaluation}
\label{subsec:model_evaluation}

\noindent \textbf{Evaluation technique}.
To evaluate these models, we randomly shuffled and split the dataset of the collected vulnerable functions into training, validation, and testing sets, with the ratios of 80:10:10, respectively.
These ratios have been widely adopted for function-level SV prediction, particularly when DL models were employed (e.g.,~\cite{li2018vuldeepecker,li2021vulnerability,hin2022linevd,fu2022linevul}), similar to our study.
We also stratified the data on all CVSS metrics to ensure that the split sets of data contained roughly the same proportion of metrics, which has been shown to be useful for defect/SV prediction~\cite{wattanakriengkrai2020predicting}.
To increase the quality of the evaluation, we also discarded the duplicate/inconsistent code functions, if any, between the sets, as per the recommendations in the literature~\cite{croft2023data}.
For RQ1, the function source code and each of the respective CVSS metrics were used as inputs for the multi-class ML and DL models, resulting in seven evaluation runs for each function.
For RQ2, all seven CVSS metrics were selected along with the function source code to be used as inputs for the multi-task models.
To find the best set of hyperparameters, grid search was used on ML models, and Ray Tune \cite{liaw2018tune} with a randomized grid search was used when evaluating the DL models.
These hyperparameters are listed in \tab~\ref{tab:hyperparameter_ml_models} and \tab~\ref{tab:hyperparameter_dl_models}. The models with the highest performance on the validation set were then selected as the optimal models, and their performance on the testing set would be reported and analyzed in RQ1 and RQ2.
We also did not use class rebalancing techniques like random over/under-sampling because they would influence SV predictive performance, making it challenging to focus on evaluating the choice of data-driven models on the performance.

\noindent \textbf{Evaluation measures}. 
Matthews Correlation Coefficient (MCC) and Macro F1-Score were used to quantify the performance of the SV assessment models.
These were widely used for SV assessment tasks (e.g.,~\mbox{\cite{spanos2018multi,kudjo2019improving,le2019automated,le2022use}}). 
F1-Score values range from 0 to 1.
MCC considers all classes with values ranging from -1 (worst) to 1 (best), not just the positive class like F1-Score. Thus, we leveraged MCC as the measure to select optimal models.

\section{Results}
\label{sec:results}

\subsection{RQ1: How well do multi-class Machine Learning and Deep Learning models perform SV assessment in C/C++?}
\label{subsec:rq1_results}
RQ1 presented and compared the performance of the multi-class ML and multi-class DL models, described in section \ref{subsec:assessment_models}, in predicting the seven CVSS metrics for vulnerable functions in C/C++.
The performance values of all of these models are reported in \tab~\ref{tab:perf}.

\noindent\textbf{Multi-class ML performance}.
The performance of ML models varied for different CVSS metrics.
The ranking of the tasks predicted by ML from high to low based on MCC values was Access Vector, Integrity, Availability, Severity, Authentication and Confidentiality, and Access Complexity.
It should be noted that while Access Vector had the highest rank based on MCC, its macro F1-Score was quite low.
Access Vector had the two majority classes, `Network' and `Local', which constituted 98.8\% of all data. Due to the extremely small amount of data, the ML models tended to produce more false-positive predictions for the minority class (`Adjacent Network').
This, in turn, lowered the macro F1-Score that considered the F1-Score values of all the classes equally.
This result also supports our choice of MCC as the main measure for choosing optimal models, as it better reflects the evaluation of imbalanced classes.

\begin{figure}[t]
    \centering
    \includegraphics[width=\columnwidth,keepaspectratio]{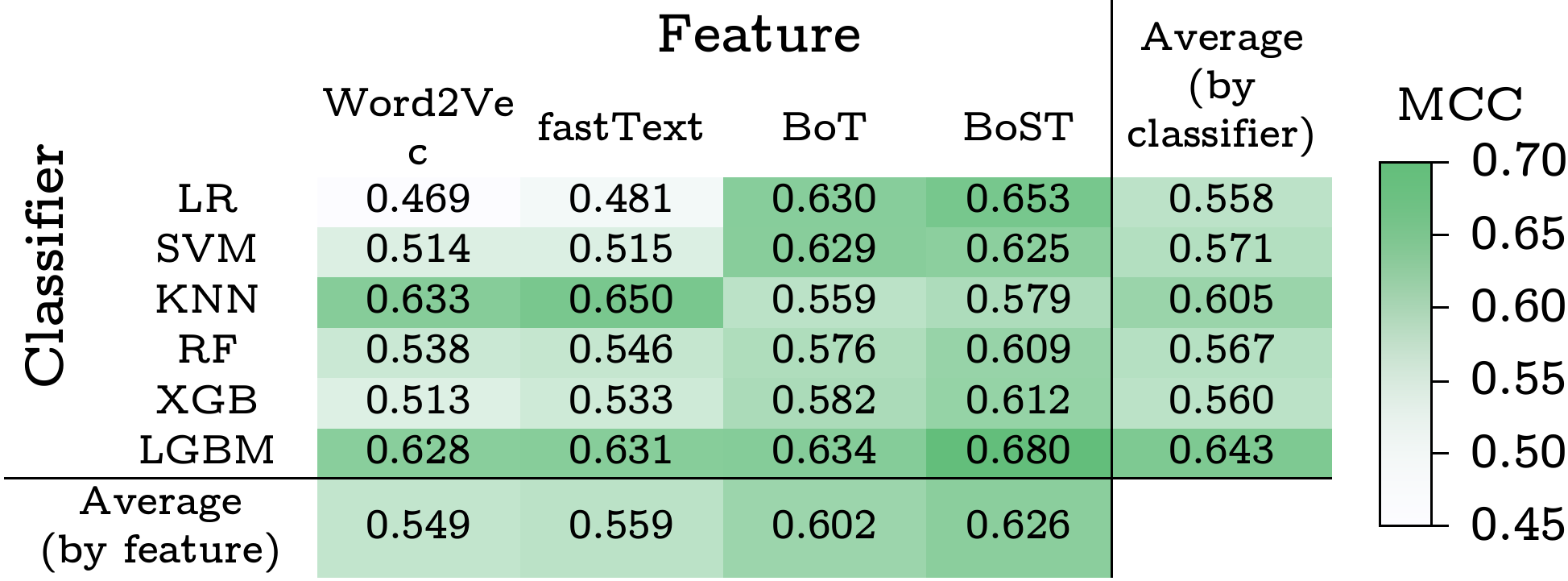}
    \caption{Average performance (MCC) of ML combinations (classifiers and features) for function-level SV assessment.}
    \label{fig:perf_ml_model_avg}
\end{figure}

We discovered several classifiers and feature extraction methods that performed consistently better than the others on average.
When averaging across all tasks as shown in \fig~\ref{fig:perf_ml_model_avg}, ensemble classifiers (RF, XGB, and LGBM) were overall better than non-ensemble ones.
Among the ensemble models, LGBM produced the best average MCC of 0.643, 15\% higher than XGB and 13\% higher than RF. 
Similarly, K-NN performed 8\% better than LR and 6\% better than SVM on average in the non-ensemble group.
In terms of feature extraction methods, features using sub-tokens generally produced higher results than their counterparts using only tokens (BoST vs. BoT and fastText vs. Word2Vec).
BoST yielded the highest average performance with MCC of 0.626, 4\% higher than BoT, while fastText is 2\% better when compared to Word2Vec.
While the MCC-wise comparisons were presented above, the F1-wise results were generally the same.
As a result, the use of LGBM with BoST produced the best overall performance (MCC: 0.680, F1: 0.763), while the LR and Word2Vec combination yielded the lowest performance (MCC: 0.469, F1: 0.614). 
The results suggest the potential effectiveness when using the specific combination of LGBM and BoST for function-level SV assessment.
Therefore, we would use LGBM + BoST for further analysis and comparison with DL models.

\noindent\textbf{Multi-class DL performance}.
Overall, we found that most DL models performed similarly, except that CodeBERT achieved the best performance, which was better than the others in terms of both MCC and F1-Score.
Among the models based on graph architectures, the GGNN models exhibited performance on par with that of the GCN counterparts.
In the non-graph category, the sequence-based models, i.e., CNN and LSTM, demonstrated similar performance.
Surpassing both sequence-based (CNN and LSTM) and graph-based models (GGNN and GCN), CodeBERT was the best-performing model with MCC and F1-Score values of 0.673 and 0.746, respectively.
The ability to generalize to tokens not found in the training data that CodeBERT~\cite{feng2020codebert} has likely allowed it to be effective in function-level SV assessment.

\noindent\textbf{Multi-class ML vs. Multi-class DL}.
The multi-class DL models showcased slightly higher average performance, 2.5\% in F1-Score and 5\% in MCC, than the ML models.
However, the best ML model (LGBM +  BoST) was actually on par with the best DL model (CodeBERT).
Access Vector was The only one of the seven tasks in which CodeBERT surpassed LGBM + BoST.
Access Vector was also the task where all DL models exhibited strong performance and outperformed the best ML model.
All the remaining tasks, i.e., Confidentiality, Integrity, Availability, Availability, and Severity, shared the same pattern in which no DL model was able to pass the best ML model (LGBM + BoST).
These findings show that the multi-class DL models demonstrated better performance for tasks with a higher degree of data imbalance, i.e., Access Vector, compared to the multi-class ML models.

In terms of efficiency, as reported in ~\fig~\ref{fig:training_time}, CodeBERT had to be trained nearly five times longer, while the overall performance was still not as effective as LGBM + BoST.
Due to their complex architectures, maintaining seven different DL models along with the tuned hyperparameters would also be much more costly than maintaining seven ML models.
In circumstances where resources are limited, multi-class DL models may not be the best choice for function-level SV assessment since they need to be trained/inferred on GPUs to maximize their capabilities, and multi-class ML models requiring only CPU will be much more efficient while still providing similar performance.
Multi-task learning can address the current challenge faced by the DL models, in which a single DL model is trained to predict all SV assessment tasks simultaneously. The results of multi-task learning for function-level SV assessment are investigated in RQ2 (see section~\ref{subsec:rq2_results}).

\begin{tcolorbox}
\textbf{RQ1 Summary}.
Regarding multi-class ML models, ensemble models beat non-ensemble ones, and sub-token features are better than token ones, resulting in LGBM + BoST being the best ML model.
Regarding multi-class DL models, CodeBERT is the best, followed by the graph-based models and finally the sequence-based models (CNN/LSTM) on average.
Overall, the best ML model, LGBM + BoST, is competitive for function-level SV assessment with comparable performance to CodeBERT and significantly (75\%) less training time.
\end{tcolorbox}

\subsection{RQ2: Can multi-task Deep Learning improve SV assessment in C++?}
\label{subsec:rq2_results}

\begin{table*}[t]
\fontsize{7}{8}\selectfont
\centering
\begin{tabular}{cc||cccccccccccccccc}
\multicolumn{2}{c||}{} & \multicolumn{2}{c|}{\textbf{\begin{tabular}[c]{@{}c@{}}Access\\ Vector\end{tabular}}} & \multicolumn{2}{c|}{\textbf{\begin{tabular}[c]{@{}c@{}}Access\\ Complexity\end{tabular}}} & \multicolumn{2}{c|}{\textbf{\begin{tabular}[c]{@{}c@{}}Authen-\\ tication\end{tabular}}} & \multicolumn{2}{c|}{\textbf{\begin{tabular}[c]{@{}c@{}}Confiden-\\ tiality\end{tabular}}} & \multicolumn{2}{c|}{\textbf{Integrity}} & \multicolumn{2}{c|}{\textbf{Availability}} & \multicolumn{2}{c|}{\textbf{Severity}} & \multicolumn{2}{c}{\textit{\begin{tabular}[c]{@{}c@{}}Average \\ (by Model)\end{tabular}}} \\ \cline{3-18} 
\multicolumn{2}{c||}{\multirow{-2}{*}{\diagbox{\textbf{Method}}{\textbf{CVSS Metric}}}} & \textbf{F1} & \multicolumn{1}{c|}{\textbf{MCC}} & \textbf{F1} & \multicolumn{1}{c|}{\textbf{MCC}} & \textbf{F1} & \multicolumn{1}{c|}{\textbf{MCC}} & \textbf{F1} & \multicolumn{1}{c|}{\textbf{MCC}} & \textbf{F1} & \multicolumn{1}{c|}{\textbf{MCC}} & \textbf{F1} & \multicolumn{1}{c|}{\textbf{MCC}} & \textbf{F1} & \multicolumn{1}{c|}{\textbf{MCC}} & \textit{F1} & \textit{MCC} \\ \hline \hline
\textbf{} & \textbf{} & \multicolumn{16}{r}{} \\
\textbf{Classifer} & \textbf{Feature} & \multicolumn{16}{c}{\multirow{-2}{*}{\textit{\textbf{Multi-Class Machine Learning Models}}}} \\ \hline \hline
 & BoT & 0.689 & \multicolumn{1}{c|}{0.789} & 0.702 & \multicolumn{1}{c|}{0.566} & 0.791 & \multicolumn{1}{c|}{0.591} & 0.746 & \multicolumn{1}{c|}{0.618} & 0.748 & \multicolumn{1}{c|}{0.621} & 0.693 & \multicolumn{1}{c|}{0.586} & 0.743 & \multicolumn{1}{c|}{0.643} & 0.730 & 0.630 \\
 & BoST & 0.740 & \multicolumn{1}{c|}{0.810} & 0.734 & \multicolumn{1}{c|}{0.574} & 0.804 & \multicolumn{1}{c|}{0.608} & 0.761 & \multicolumn{1}{c|}{0.631} & 0.766 & \multicolumn{1}{c|}{0.653} & 0.726 & \multicolumn{1}{c|}{0.638} & \cellcolor[HTML]{AEAAAA}0.759 & \multicolumn{1}{c|}{0.656} & 0.756 & 0.653 \\
 & Word2Vec & 0.506 & \multicolumn{1}{c|}{0.415} & 0.554 & \multicolumn{1}{c|}{0.415} & 0.782 & \multicolumn{1}{c|}{0.592} & 0.628 & \multicolumn{1}{c|}{0.439} & 0.676 & \multicolumn{1}{c|}{0.505} & 0.573 & \multicolumn{1}{c|}{0.470} & 0.577 & \multicolumn{1}{c|}{0.447} & 0.614 & 0.469 \\
\multirow{-4}{*}{LR} & fastText & 0.504 & \multicolumn{1}{c|}{0.416} & 0.580 & \multicolumn{1}{c|}{0.416} & 0.783 & \multicolumn{1}{c|}{0.629} & 0.638 & \multicolumn{1}{c|}{0.451} & 0.668 & \multicolumn{1}{c|}{0.501} & 0.597 & \multicolumn{1}{c|}{0.481} & 0.595 & \multicolumn{1}{c|}{0.477} & 0.623 & 0.481 \\ \hline
 & BoT & 0.716 & \multicolumn{1}{c|}{0.784} & 0.709 & \multicolumn{1}{c|}{0.564} & 0.811 & \multicolumn{1}{c|}{0.629} & 0.741 & \multicolumn{1}{c|}{0.611} & 0.738 & \multicolumn{1}{c|}{0.611} & 0.693 & \multicolumn{1}{c|}{0.585} & 0.727 & \multicolumn{1}{c|}{0.623} & 0.734 & 0.629 \\
 & BoST & 0.724 & \multicolumn{1}{c|}{0.787} & 0.700 & \multicolumn{1}{c|}{0.555} & 0.753 & \multicolumn{1}{c|}{0.511} & 0.748 & \multicolumn{1}{c|}{0.615} & 0.757 & \multicolumn{1}{c|}{0.634} & 0.720 & \multicolumn{1}{c|}{0.622} & 0.756 & \multicolumn{1}{c|}{0.648} & 0.737 & 0.625 \\
 & Word2Vec & 0.632 & \multicolumn{1}{c|}{0.730} & 0.561 & \multicolumn{1}{c|}{0.419} & 0.773 & \multicolumn{1}{c|}{0.592} & 0.629 & \multicolumn{1}{c|}{0.441} & 0.668 & \multicolumn{1}{c|}{0.503} & 0.570 & \multicolumn{1}{c|}{0.465} & 0.563 & \multicolumn{1}{c|}{0.447} & 0.628 & 0.514 \\
\multirow{-4}{*}{SVM} & fastText & 0.579 & \multicolumn{1}{c|}{0.724} & 0.546 & \multicolumn{1}{c|}{0.415} & 0.727 & \multicolumn{1}{c|}{0.544} & 0.636 & \multicolumn{1}{c|}{0.447} & 0.669 & \multicolumn{1}{c|}{0.504} & 0.589 & \multicolumn{1}{c|}{0.491} & 0.594 & \multicolumn{1}{c|}{0.479} & 0.620 & 0.515 \\ \hline
 & BoT & 0.634 & \multicolumn{1}{c|}{0.730} & 0.650 & \multicolumn{1}{c|}{0.455} & 0.756 & \multicolumn{1}{c|}{0.588} & 0.679 & \multicolumn{1}{c|}{0.521} & 0.706 & \multicolumn{1}{c|}{0.560} & 0.658 & \multicolumn{1}{c|}{0.544} & 0.658 & \multicolumn{1}{c|}{0.517} & 0.677 & 0.559 \\
 & BoST & 0.650 & \multicolumn{1}{c|}{0.709} & 0.683 & \multicolumn{1}{c|}{0.478} & 0.787 & \multicolumn{1}{c|}{0.601} & 0.715 & \multicolumn{1}{c|}{0.571} & 0.721 & \multicolumn{1}{c|}{0.581} & 0.680 & \multicolumn{1}{c|}{0.582} & 0.672 & \multicolumn{1}{c|}{0.530} & 0.701 & 0.579 \\
 & Word2Vec & 0.714 & \multicolumn{1}{c|}{0.800} & 0.719 & \multicolumn{1}{c|}{0.582} & 0.763 & \multicolumn{1}{c|}{0.560} & 0.745 & \multicolumn{1}{c|}{0.612} & 0.745 & \multicolumn{1}{c|}{0.618} & 0.729 & \multicolumn{1}{c|}{0.635} & 0.731 & \multicolumn{1}{c|}{0.624} & 0.735 & 0.633 \\
\multirow{-4}{*}{K-NN} & fastText & 0.727 & \multicolumn{1}{c|}{0.795} & \cellcolor[HTML]{AEAAAA}0.760 & \multicolumn{1}{c|}{\cellcolor[HTML]{AEAAAA}0.626} & 0.763 & \multicolumn{1}{c|}{0.560} & 0.751 & \multicolumn{1}{c|}{0.619} & 0.772 & \multicolumn{1}{c|}{0.658} & \cellcolor[HTML]{AEAAAA}0.738 & \multicolumn{1}{c|}{0.649} & 0.735 & \multicolumn{1}{c|}{0.642} & 0.749 & 0.650 \\ \hline
 & BoT & 0.587 & \multicolumn{1}{c|}{0.749} & 0.531 & \multicolumn{1}{c|}{0.481} & 0.808 & \multicolumn{1}{c|}{0.667} & 0.623 & \multicolumn{1}{c|}{0.458} & 0.673 & \multicolumn{1}{c|}{0.553} & 0.599 & \multicolumn{1}{c|}{0.561} & 0.594 & \multicolumn{1}{c|}{0.567} & 0.631 & 0.576 \\
 & BoST & 0.638 & \multicolumn{1}{c|}{0.750} & 0.602 & \multicolumn{1}{c|}{0.516} & \multicolumn{1}{c}{\cellcolor[HTML]{AEAAAA}0.853} & \multicolumn{1}{c|}{\cellcolor[HTML]{AEAAAA}0.738} & 0.660 & \multicolumn{1}{c|}{0.502} & 0.706 & \multicolumn{1}{c|}{0.591} & 0.634 & \multicolumn{1}{c|}{0.582} & 0.606 & \multicolumn{1}{c|}{0.584} & 0.671 & 0.609 \\
 & Word2Vec & 0.634 & \multicolumn{1}{c|}{0.730} & 0.500 & \multicolumn{1}{c|}{0.456} & 0.626 & \multicolumn{1}{c|}{0.384} & 0.674 & \multicolumn{1}{c|}{0.530} & 0.707 & \multicolumn{1}{c|}{0.577} & 0.609 & \multicolumn{1}{c|}{0.562} & 0.561 & \multicolumn{1}{c|}{0.531} & 0.616 & 0.538 \\
\multirow{-4}{*}{RF} & fastText & 0.635 & \multicolumn{1}{c|}{0.742} & 0.508 & \multicolumn{1}{c|}{0.480} & 0.626 & \multicolumn{1}{c|}{0.384} & 0.676 & \multicolumn{1}{c|}{0.522} & 0.705 & \multicolumn{1}{c|}{0.576} & 0.606 & \multicolumn{1}{c|}{0.554} & 0.588 & \multicolumn{1}{c|}{0.564} & 0.621 & 0.546 \\ \hline
 & BoT & 0.639 & \multicolumn{1}{c|}{0.750} & 0.625 & \multicolumn{1}{c|}{0.479} & 0.797 & \multicolumn{1}{c|}{0.632} & 0.669 & \multicolumn{1}{c|}{0.499} & 0.703 & \multicolumn{1}{c|}{0.572} & 0.635 & \multicolumn{1}{c|}{0.562} & 0.658 & \multicolumn{1}{c|}{0.578} & 0.675 & 0.582 \\
 & BoST & 0.711 & \multicolumn{1}{c|}{0.772} & 0.662 & \multicolumn{1}{c|}{0.531} & 0.820 & \multicolumn{1}{c|}{0.670} & 0.686 & \multicolumn{1}{c|}{0.528} & 0.736 & \multicolumn{1}{c|}{0.606} & 0.670 & \multicolumn{1}{c|}{0.601} & 0.673 & \multicolumn{1}{c|}{0.572} & 0.708 & 0.612 \\
 & Word2Vec & 0.710 & \multicolumn{1}{c|}{0.750} & 0.568 & \multicolumn{1}{c|}{0.419} & 0.718 & \multicolumn{1}{c|}{0.502} & 0.653 & \multicolumn{1}{c|}{0.476} & 0.655 & \multicolumn{1}{c|}{0.488} & 0.611 & \multicolumn{1}{c|}{0.510} & 0.578 & \multicolumn{1}{c|}{0.445} & 0.642 & 0.513 \\
\multirow{-4}{*}{XGB} & fastText & 0.667 & \multicolumn{1}{c|}{0.757} & 0.599 & \multicolumn{1}{c|}{0.456} & 0.696 & \multicolumn{1}{c|}{0.496} & 0.652 & \multicolumn{1}{c|}{0.470} & 0.678 & \multicolumn{1}{c|}{0.531} & 0.615 & \multicolumn{1}{c|}{0.522} & 0.611 & \multicolumn{1}{c|}{0.500} & 0.645 & 0.533 \\ \hline
 & BoT & 0.732 & \multicolumn{1}{c|}{0.789} & 0.687 & \multicolumn{1}{c|}{0.563} & 0.727 & \multicolumn{1}{c|}{0.544} & 0.762 & \multicolumn{1}{c|}{0.630} & 0.755 & \multicolumn{1}{c|}{0.640} & 0.706 & \multicolumn{1}{c|}{0.630} & 0.734 & \multicolumn{1}{c|}{0.639} & 0.729 & 0.634 \\
 & BoST & {\cellcolor[HTML]{AEAAAA}\color[HTML]{3166FF} 0.754} & \multicolumn{1}{c|}{{\color[HTML]{3166FF} 0.803}} & {\color[HTML]{3166FF} 0.732} & \multicolumn{1}{c|}{{\color[HTML]{3166FF} 0.609}} & {\color[HTML]{3166FF} 0.783} & \multicolumn{1}{c|}{{\color[HTML]{3166FF} 0.629}} & \cellcolor[HTML]{AEAAAA}{\color[HTML]{3166FF} 0.780} & \multicolumn{1}{c|}{\cellcolor[HTML]{AEAAAA}{\color[HTML]{3166FF} 0.659}} & \cellcolor[HTML]{AEAAAA}{\color[HTML]{3166FF} 0.801} & \multicolumn{1}{c|}{\cellcolor[HTML]{AEAAAA}{\color[HTML]{3166FF} 0.703}} & {\color[HTML]{3166FF} 0.737} & \multicolumn{1}{c|}{\cellcolor[HTML]{AEAAAA}{\color[HTML]{3166FF} 0.678}} & {\color[HTML]{3166FF} 0.751} & \multicolumn{1}{c|}{\cellcolor[HTML]{AEAAAA}{\color[HTML]{3166FF} 0.680}} & {\cellcolor[HTML]{AEAAAA}\color[HTML]{3166FF} 0.763} & \cellcolor[HTML]{AEAAAA}{\color[HTML]{3166FF} 0.680} \\
 & Word2Vec & 0.702 & \multicolumn{1}{c|}{\cellcolor[HTML]{AEAAAA}0.811} & 0.695 & \multicolumn{1}{c|}{0.585} & 0.718 & \multicolumn{1}{c|}{0.502} & 0.737 & \multicolumn{1}{c|}{0.599} & 0.750 & \multicolumn{1}{c|}{0.629} & 0.701 & \multicolumn{1}{c|}{0.624} & 0.729 & \multicolumn{1}{c|}{0.650} & 0.719 & 0.628 \\
\multirow{-4}{*}{LGBM} & fastText & 0.697 & \multicolumn{1}{c|}{0.795} & 0.694 & \multicolumn{1}{c|}{0.598} & 0.688 & \multicolumn{1}{c|}{0.451} & 0.743 & \multicolumn{1}{c|}{0.603} & 0.770 & \multicolumn{1}{c|}{0.659} & 0.723 & \multicolumn{1}{c|}{0.653} & 0.737 & \multicolumn{1}{c|}{0.659} & 0.722 & 0.631 \\ \hline
\multicolumn{2}{c||}{\textit{\begin{tabular}[c]{@{}c@{}}Average (by\\ CVSS Metric)\end{tabular}}} & 0.670 & \multicolumn{1}{c|}{0.737} & 0.638 & \multicolumn{1}{c|}{0.510} & 0.756 & \multicolumn{1}{c|}{0.567} & 0.697 & \multicolumn{1}{c|}{0.544} & 0.720 & \multicolumn{1}{c|}{0.586} & 0.659 & \multicolumn{1}{c|}{0.574} & 0.664 & \multicolumn{1}{c|}{0.571} & 0.687 & 0.584 \\ \hline \hline

\multicolumn{2}{c||}{} & \multicolumn{16}{r}{} \\
\textbf{Type} & \multicolumn{1}{c||}{\textbf{Model}} & \multicolumn{16}{c}{\multirow{-2}{*}{\textit{\textbf{Multi-Class Deep Learning Models}}}} \\ \hline \hline
 & GGNN & 0.765 & \multicolumn{1}{c|}{0.816} & 0.716 & \multicolumn{1}{c|}{0.548} & 0.783 & \multicolumn{1}{c|}{0.591} & 0.674 & \multicolumn{1}{c|}{0.518} & 0.739 & \multicolumn{1}{c|}{0.618} & 0.719 & \multicolumn{1}{c|}{0.606} & 0.694 & \multicolumn{1}{c|}{0.578} & 0.727 & 0.611 \\
\multirow{-2}{*}{\begin{tabular}[c]{@{}c@{}}Graph-\\ based\end{tabular}} & GCN & \cellcolor[HTML]{AEAAAA}0.826 & \multicolumn{1}{c|}{0.808} & 0.734 & \multicolumn{1}{c|}{0.594} & 0.785 & \multicolumn{1}{c|}{0.572} & 0.738 & \multicolumn{1}{c|}{0.606} & 0.717 & \multicolumn{1}{c|}{0.579} & 0.681 & \multicolumn{1}{c|}{0.551} & 0.671 & \multicolumn{1}{c|}{0.549} & 0.736 & 0.608 \\ \hline
 & CNN & 0.619 & \multicolumn{1}{c|}{0.844} & 0.484 & \multicolumn{1}{c|}{0.452} & 0.763 & \multicolumn{1}{c|}{0.538} & 0.728 & \multicolumn{1}{c|}{0.595} & \cellcolor[HTML]{AEAAAA}0.775 & \multicolumn{1}{c|}{\cellcolor[HTML]{AEAAAA}0.669} & 0.754 & \multicolumn{1}{c|}{0.670} & 0.451 & \multicolumn{1}{c|}{0.407} & 0.653 & 0.596 \\
& LSTM & 0.619 & \multicolumn{1}{c|}{0.839} & 0.731 & \multicolumn{1}{c|}{\cellcolor[HTML]{AEAAAA}0.626} & 0.729 & \multicolumn{1}{c|}{0.469} & 0.423 & \multicolumn{1}{c|}{0.337} & 0.768 & \multicolumn{1}{c|}{0.658} & 0.707 & \multicolumn{1}{c|}{0.623} & 0.745 & \multicolumn{1}{c|}{0.653} & 0.675 & 0.601 \\
\multirow{-3}{*}{\begin{tabular}[c]{@{}c@{}}Non-\\ Graph\end{tabular}} & CodeBERT & 0.619 & \multicolumn{1}{c|}{\cellcolor[HTML]{AEAAAA}0.845} & \cellcolor[HTML]{AEAAAA}0.739 & \multicolumn{1}{c|}{0.618} & \multicolumn{1}{c}{\cellcolor[HTML]{AEAAAA}0.800} & \multicolumn{1}{c|}{\cellcolor[HTML]{AEAAAA}0.602} & \cellcolor[HTML]{AEAAAA}0.765 & \multicolumn{1}{c|}{\cellcolor[HTML]{AEAAAA}0.649} & 0.749 & \multicolumn{1}{c|}{0.637} & \cellcolor[HTML]{AEAAAA}0.763 & \multicolumn{1}{c|}{\cellcolor[HTML]{AEAAAA}0.691} & \cellcolor[HTML]{AEAAAA}0.786 & \multicolumn{1}{c|}{\cellcolor[HTML]{AEAAAA}0.669} & \cellcolor[HTML]{AEAAAA}0.746 & \cellcolor[HTML]{AEAAAA}0.673 \\ \hline
\multicolumn{2}{c||}{\textit{\begin{tabular}[c]{@{}c@{}}Average (by\\ CVSS Metric)\end{tabular}}} & \textit{0.683} & \multicolumn{1}{c|}{\textit{0.825}} & \textit{0.671} & \multicolumn{1}{c|}{\textit{0.558}} & \textit{0.780} & \multicolumn{1}{c|}{\textit{0.573}} & \textit{0.673} & \multicolumn{1}{c|}{\textit{0.545}} & \textit{0.743} & \multicolumn{1}{c|}{\textit{0.623}} & \textit{0.712} & \multicolumn{1}{c|}{\textit{0.611}} & \textit{0.667} & \multicolumn{1}{c|}{\textit{0.559}} & \textit{0.704} & \textit{0.613} \\ \hline \hline

\multicolumn{2}{c||}{} & \multicolumn{16}{r}{} \\
\textbf{Type} & \multicolumn{1}{c||}{\textbf{Model}} & \multicolumn{16}{c}{\multirow{-2}{*}{\textit{\textbf{Multi-Task Deep Learning Models}}}} \\ \hline \hline
 & GGNN & 0.700 & \multicolumn{1}{c|}{0.828} & 0.688 & \multicolumn{1}{c|}{0.831} & 0.831 & \multicolumn{1}{c|}{0.676} & 0.756 & \multicolumn{1}{c|}{0.627} & 0.760 & \multicolumn{1}{c|}{0.636} & 0.725 & \multicolumn{1}{c|}{0.645} & 0.727 & \multicolumn{1}{c|}{0.627} & 0.741 & 0.696 \\
\multirow{-2}{*}{\begin{tabular}[c]{@{}c@{}}Graph-\\ based\end{tabular}} & GCN & 0.614 & \multicolumn{1}{c|}{\cellcolor[HTML]{AEAAAA}0.874} & 0.764 & \multicolumn{1}{c|}{0.831} & 0.831 & \multicolumn{1}{c|}{0.664} & 0.750 & \multicolumn{1}{c|}{\cellcolor[HTML]{AEAAAA}0.674} & 0.753 & \multicolumn{1}{c|}{0.642} & 0.750 & \multicolumn{1}{c|}{0.666} & 0.728 & \multicolumn{1}{c|}{0.616} & 0.741 & 0.709 \\ \hline
  & CNN & 0.734 & \multicolumn{1}{c|}{0.853} & 0.734 & \multicolumn{1}{c|}{\cellcolor[HTML]{AEAAAA}0.843} & \cellcolor[HTML]{AEAAAA}0.843 & \multicolumn{1}{c|}{\cellcolor[HTML]{AEAAAA}0.707} & 0.766 & \multicolumn{1}{c|}{0.651} & 0.762 & \multicolumn{1}{c|}{0.652} & 0.793 & \multicolumn{1}{c|}{0.708} & 0.772 & \multicolumn{1}{c|}{0.677} & 0.772 & 0.727 \\
 & LSTM & 0.617 & \multicolumn{1}{c|}{0.840} & 0.677 & \multicolumn{1}{c|}{0.805} & 0.805 & \multicolumn{1}{c|}{0.663} & 0.733 & \multicolumn{1}{c|}{0.599} & 0.756 & \multicolumn{1}{c|}{0.639} & 0.710 & \multicolumn{1}{c|}{0.605} & 0.740 & \multicolumn{1}{c|}{0.634} & 0.720 & 0.683 \\
    \multirow{-3}{*}{\begin{tabular}[c]{@{}c@{}}Non-\\ Graph\end{tabular}} & CodeBERT & \cellcolor[HTML]{AEAAAA}0.851 & \multicolumn{1}{c|}{0.857} & \cellcolor[HTML]{AEAAAA}0.785 & \multicolumn{1}{c|}{0.840} & 0.841 & \multicolumn{1}{c|}{0.680} & \cellcolor[HTML]{AEAAAA}0.799 & \multicolumn{1}{c|}{0.650} & \cellcolor[HTML]{AEAAAA}0.802 & \multicolumn{1}{c|}{\cellcolor[HTML]{AEAAAA}0.684} & \cellcolor[HTML]{AEAAAA}0.817 & \multicolumn{1}{c|}{\cellcolor[HTML]{AEAAAA}0.711} & \cellcolor[HTML]{AEAAAA}0.845 & \multicolumn{1}{c|}{\cellcolor[HTML]{AEAAAA}0.715} & \cellcolor[HTML]{AEAAAA}0.820 & \cellcolor[HTML]{AEAAAA}0.734 \\ \hline
\multicolumn{2}{c||}{\textit{\begin{tabular}[c]{@{}c@{}}Average (by\\ CVSS Metric)\end{tabular}}} & \textit{0.727} & \multicolumn{1}{c|}{\textit{0.846}} & \textit{0.741} & \multicolumn{1}{c|}{\textit{0.827}} & \textit{0.828} & \multicolumn{1}{c|}{\textit{0.669}} & \textit{0.767} & \multicolumn{1}{c|}{\textit{0.638}} & \textit{0.773} & \multicolumn{1}{c|}{\textit{0.654}} & \textit{0.762} & \multicolumn{1}{c|}{\textit{0.672}} & \textit{0.770} & \multicolumn{1}{c|}{\textit{0.655}} & \textit{0.759} & \textit{0.709} \\ \hline
\end{tabular}
\caption{Testing performance of different Machine Learning and Deep Learning models for CVSS-based function-level SV assessment. Notes: Values highlighted in grey are best task-wise for each method. Values in blue belong to the best ML model, i.e., LGBM + BoST.}
\label{tab:perf}
\end{table*}

Instead of developing multi-class models for each of the seven CVSS metrics as in RQ1, in RQ2, we built unified DL models with multi-task learning (see~\ref{subsec:assessment_dl_models}). In this scenario, each model predicted all the function-level SV assessment outputs in a single model.
This setting is expected to improve the overall effectiveness and efficiency of the DL models. We compared the multi-task performance with that of the best multi-class models in RQ1 (see \tab~\ref{tab:perf}).

\begin{figure}[t]
    \centering
    \includegraphics[width=\columnwidth,keepaspectratio]{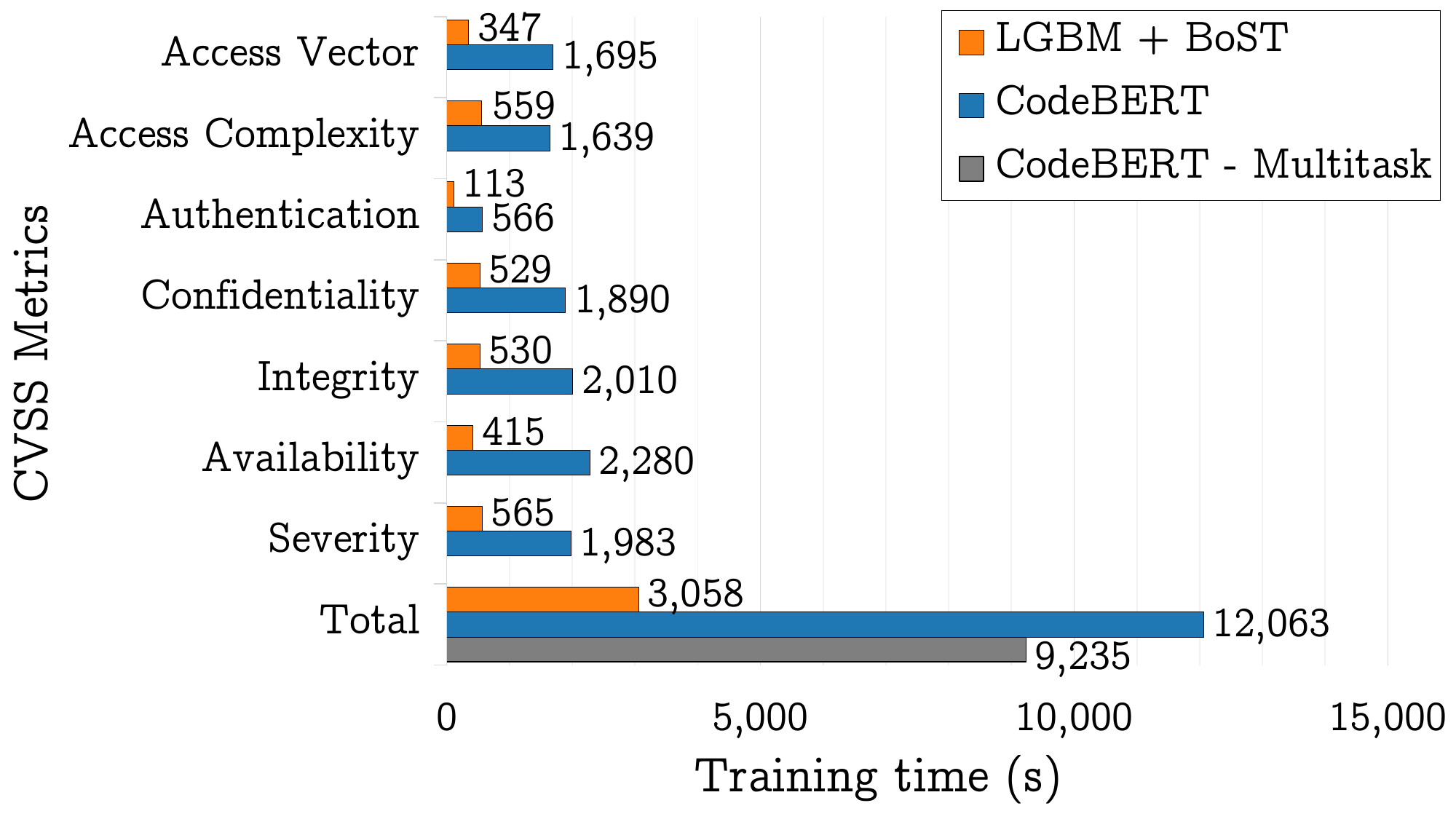}
    \caption{Training time of CodeBERT, multi-task CodeBERT and LGBM + BoST (in seconds). Note: CodeBERT multi-task does not have training time for each task.}
    \label{fig:training_time}
    \vspace{-10pt}
\end{figure}

In the multi-task training scenario, CodeBERT was still the most effective model, outperforming the best ML model (LGBM + BoST) by 8\% in both MCC and F1-Score.
For individual tasks, the multi-task CodeBERT variant outperformed LGBM + BoST in five tasks: Access Vector, Access Complexity, Authentication, Availability, and Severity, with MCC increases of 7\%, 38\%, 8\%, 5\%, and 5\%, respectively.
For the remaining Confidentiality and Integrity tasks, the multi-task CodeBERT model still performed reasonably well within 3\% of the MCC value of the ML model.
These improvements imply that multi-task learning is particularly useful for the tasks with more imbalanced data, i.e., Access Vector, Access Complexity, and Authentication.
It is worth noting that similar to RQ1, the multi-task variants of CNN, LSTM, GGNN, and GCN attained similar levels of performance for all the SV assessment tasks.

Further, multi-task learning was beneficial to all sequence-based, graph-based, and Transformer-based DL models in terms of both MCC and F1-Score (see \tab~\ref{tab:perf}). 
MCC improved significantly at 22\%, 14\%, 14\%, 17\%, and 9\%, for CNN, LSTM, GGNN, GCN, and CodeBERT, respectively.
The multi-task implementation also demonstrated better MCC values than the multi-class counterparts across all the tasks.
Specifically, the improvements were 3\%, 48\%, 17\%, 17\%, 5\%, 10\%, and 17\% for Access Vector, Access Complexity, Authentication, Confidentiality, Integrity, Availability, and Severity, respectively.
The improvement for Access Vector was the least, probably because DL models already performed best for this task, as shown in RQ1.
It is also important to note that the best multi-task CodeBERT model cut down the total training time of the multi-class CodeBERT model in RQ1 by nearly 23\%, as illustrated in \fig~\ref{fig:training_time}.
This result highlights the better efficiency of multi-task learning than training seven separate multi-class DL models. Note that ML is still more efficient, yet not performing as well, when compared to multi-task learning.

\begin{tcolorbox}[before skip balanced=15pt]
\textbf{RQ2 Summary}.
Multi-task DL models achieve performance improvements of 9--22\% in MCC compared to their multi-class counterparts.
Multi-task CodeBERT is the best model overall and outperforms the best ML model (LGBM + BoST) by 8\% in MCC averaging across the tasks.
The findings suggest that multi-task learning can be employed to increase both effectiveness and efficiency.
\end{tcolorbox}

\section{Discussion}
\label{sec:discussion}

\subsection{Comparison with the literature and beyond}
\noindent\textbf{ML performance}. 
Our findings in RQ1 that sub-tokens performed better than tokens alone and ensemble classifiers were superior to non-ensemble ones align with the previous findings of SV assessment using SV reports~\cite{spanos2018multi,le2019automated}.
In addition, given that Le et al.~\cite{le2022use} also used ML for function-level SV assessment in Java (not C/C++), we also compare our findings with theirs.
We found similar results to the previous study~\cite{le2022use}, where LGBM and BoST are the best overall combination to assess SVs at the function level.
This shows the possibility of reusing the knowledge and the ML model developed for one language to another language for function-level SV assessment.
However, the ranking of the seven CVSS metrics predicted by ML, in terms of MCC, in the previous study did not exactly match ours. Among the CVSS metrics, the rank of Authentication was changed the most significantly, i.e., from rank 2 in the previous study to rank 6 in our study.
This is likely due to distinctive input code structures and semantics (C/C++ vs. Java) and different output class distributions fed into the models.
For example, for Authentication, the decrease in the rank is likely due to the much smaller proportion (2\%) of the minority class compared to that (21.8\%) in the previous study.
Such differences reinforce the need to (re-)evaluate existing models in a different language than the one for which they were originally experimented.

\noindent\textbf{DL performance}.
Fu et al. \cite{fu2022linevul} compared CodeBERT with graph-based models, including GGNN and GCN, for function-level SV prediction and showed that CodeBERT was the best overall model. Zhou et al. \cite{zhou2019devign} and Nguyen et al. \cite{nguyen2022regvd} also showed that graph-based models performed better than convolutional and recurrent models with a noticeable difference. All of these agree with our findings in RQ1 and RQ2. In brief, the ranking in decreasing performance for C/C++ function level SV assessment are as follows: Transformer-based models (CodeBERT) > Graph-based models (GGNN and GCN) > Sequence-based models (CNN and LSTM).
However, these prior studies have not investigated the use of multi-task learning as the output nature of SV detection and SV assessment is different.
Our study shows that using DL models with multi-task learning is more suitable than the conventional multi-class approach for function-level SV assessment in C/C++, improving both the effectiveness and efficiency of the tasks.

\noindent\textbf{ML versus DL}. While previous studies have shown that ML can be inferior to DL for SV prediction, this does not necessarily translate into function-level SV assessment in C/C++.
For instance, Fu et al.~\cite{fu2022linevul} claimed that RF + BoW was much less effective than CodeBERT for function-level SV detection.
While our experiments confirmed the previous finding that RF + BoW was worse than CodeBERT, we still found some ML models that performed competitively compared to DL.
Particularly, in the same multi-class setting, we showed that LGBM + BoST performed similarly to CodeBERT in both F1-Score (0.763 versus 0.746) and MCC (0.680 versus 0.673), while requiring only 1/4 of the training time.
These findings actually echo the recent recommendations in the field of Software Engineering that a carefully designed and tuned ML model can still outperform more advanced DL models while saving a significant amount of cost (e.g.,~\cite{majumder2018500+,xu2018prediction}).
For function-level SV detection, the demonstrated promising results can inspire future research to consider LGBM + BoST instead of the traditional RF + BoW as a baseline to compare with more sophisticated DL models proposed for the task.
For function-level SV assessment, multi-task DL models are still preferred if researchers or practitioners want to optimize (baseline) performance for testing/comparing with newly proposed SV assessment models/techniques.
On the other hand, LGBM + BoST can still be used to achieve reasonable performance with much fewer resources. This is particularly useful when practitioners need to routinely retrain the models in continuous development environments with frequent software releases~\cite{arani2024systematic}.

\subsection{Threats to validity}
The first threat concerns the implementation of existing models. We ensured the robustness of our implementation by following the instructions given in previous studies and only reused the validated and reviewed reproduction packages.

The second threat pertains to our conversion of the multi-class DL models to the multi-task variants. By only changing the output layers and the loss functions to accommodate the new outputs in each model, we ensured the validity of the original models, while allowing us to examine their performance without affecting the original architectures.

The third threat involves the optimization of the models. It is widely known that it is not practically possible to experiment with all possible combinations of hyperparameters. Instead, we relied on related work and focused on the options that had been demonstrated to be suitable and effective for SV-related tasks.

Another threat is the generalizability of our study.
Our findings may not generalize to all the projects in C/C++; however, our dataset comes from an established source \cite{fan2020ac}, and the code functions are derived from 300+ real projects in a variety of applications and domains.
The extensiveness and diversity of the dataset also allow for robust analysis, reducing the randomness in our results.

\section{Conclusions and Future Work}
\label{sec:conclusions}
We demonstrated the need and potential for using data-driven approaches for function-level SV assessment to streamline the SV fixing process in C/C++.
Using a customized SV dataset in C/C++, we investigated and compared the performance of various ML and DL models to predict seven CVSS metrics to assess SVs in code functions in these languages. 
When building separate models for individual tasks, we showed that ML was a competitive approach to DL for function-level SV assessment with comparable performance and less training time.
We also illustrated a better way to use DL models through multi-task learning for SV assessment tasks, which presented improvements of 8--22\% in MCC over the multi-class counterparts.
Our findings give recommendations on what and how ML and DL models can be used for function-level SV assessment.
We also show the possibility of reusing data-driven models between SV assessment and SV detection, which can draw better synergies between the two important tasks.
Overall, our work provides data and competitive models that can be used for future research to benchmark next data-driven advances for automating SV assessment in particular and SV management in general.

\balance

\bibliographystyle{ACM-Reference-Format}
\bibliography{reference}

\end{document}